# Analysis of GPU Parallel Computing based on Matlab

Mingzhe Wang, Bo Wang, Qiu He, Xiuxiu Liu, Kunshuai Zhu

(School of Computer and Control Engineering, University of Chinese Academy of Sciences, Huairou, Beijing 101408, China)

**Abstract:** Matlab is very widely used in scientific computing, but Matlab computational efficiency is lower than C language program. In order to improve the computing speed, some toolbox can use GPU to accelerate the computation. This paper describes GPU working principle, our experiments and results analysis of parallel computing by using GPU based on Matlab. Experimental results show that for parallel operations, GPU computing speed is faster than CPU, for the logical instructions, GPU computing speed is slower than CPU.

## 1. Introduction

In scientific computing, Matlab is a visualization software which contains numerical analysis, matrix operations, signal processing and graphical display. It contains rich toolbox functions, and can get a good solution of problems in the field of system simulation and calculation which encountered in the study. But Matlab computing efficiency is low, compared with other high-level languages, Matlab program execution is slow[1].

Scientific researchers always use two measures to improve the speed of Matlab. One is to buy some expensive equipments such as servers or workstations to increase hardware performance, although this method is able to solve the speed problem, but it gives researchers a financial burden[1]. The other way is transplanting MATLAB algorithms and reprogramming by high level languages such as C++, which can improve efficiency of computing, but it demand programming skills for researchers, and some algorithms are complex, high-level languages such as C++ don't provide the appropriate library functions, this requires researchers to start at the bottom. It would be very time-consuming and laborious, and increase research effort[1].

In order to solve this problem, MathWorks announced that by using Parallel Computing Toolbox, MATLAB can provide support for NVIDIA GPU. This support will allow engineers and scientists make MATLAB computing to be faster, and without having to do reprogramming, or increase equipment cost.

Our work is to conduct a series of experiments, such as FFT, matrix multiplication,

quicksort, simulate encoding and decoding Hamming code in BSC channel to compare the advantages and disadvantages in scientific computing of CPU and GPU. Results show that for high parallel operations, GPU computing speed is faster; for the logical instructions, GPU computing speed is slower. And improving the size of computing can enhance the efficiency of GPU.

## 2. Architecture of GPU

General purpose computation by GPU has been a trend of computer science. GPU is called "Graphic Processing Unit". Many researchers have noted the potential of GPU to calculate, in 2003, at the SIGGRAPH conference, researchers presented GPGPU (General-purpose computing on graphics processing units) concept[2].

### 2.1 Comparison between GPU and CPU

Relative to CPU, GPU has a longer pipeline, and doesn't have a cache[2]. GPU use more transistors as performing units, rather than as a complex system like cache and control units in CPU which are in order to improve the efficiency of a small amount of execution units[2]. Figure 1 compared the logical schema of CPU and GPU.

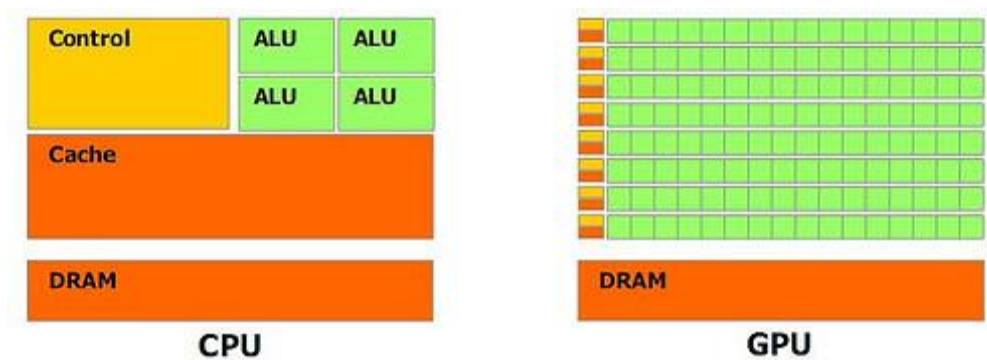

**Figure 1.** Logical schema of CPU and GPU.

Integer computations, branch instructions, logic and floating point operations of CPU are implemented by different operational units, therefore, with different types of computing tasks, CPU have different performance. While the GPU is use the same unit to perform integer and floating point calculations, so the integer and floating-point computing performance is similar.

### 2.2 Advantage of GPU

At present, mainstream GPU uses the unified architecture and units, with a powerful lineup of programmable stream processors, GPU performance of single precision floating-point is much faster than CPU[3]. Figure 2 shows the floating-point operation speed of CPU and GPU.

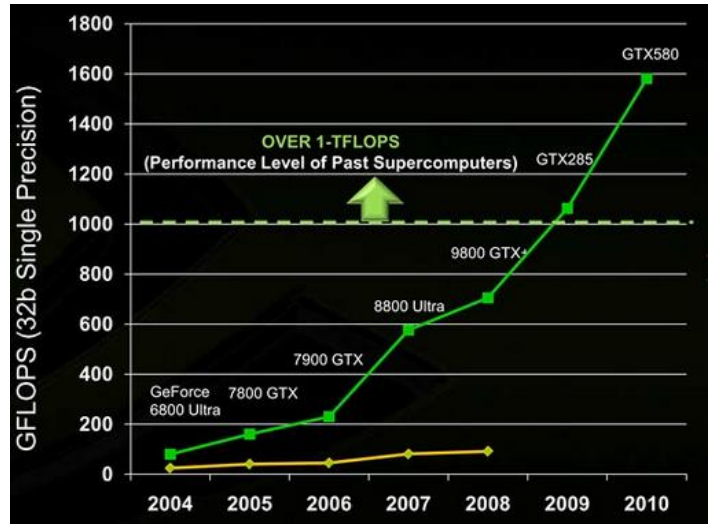

**Figure 2.** Floating-point operation speed of CPU and GPU.

GPU also has a huge advantage, which is its memory subsystem[2]. Figure 3 shows the memory bandwidth of CPU and GPU. Current common DDR3-1333 memory bandwidth is 32GB/S, but mainstream GPU has a 40-60 GB/s memory bandwidth, NVIDIA GTX280's bandwidth is 142GB/s[2]. Ultra high memory bandwidth make floating-point computation can throughput stable.

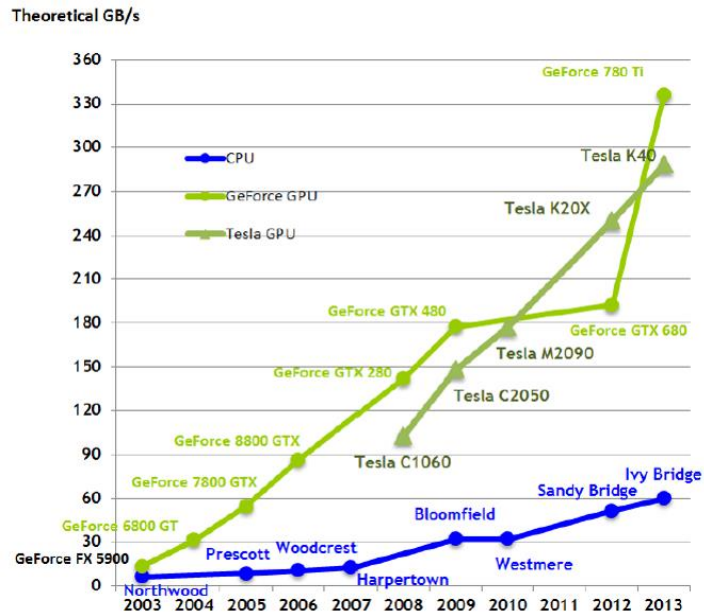

**Figure 3.** Memory bandwidth of CPU and GPU.

## 2.3 Features of suitable program for GPU computing

Although GPU computing is fast, but the GPU cannot completely replace the CPU, many operating systems, software, and code cannot run on the GPU. Program for GPU computing has following four characteristics: computer-intensive, highly parallel, simple operation, multiple stages of computing[3].

# 3. The proposed experiment

We use Parallel Computing Toolbox in Matlab, by following these four experiments, we analyze the advantages and disadvantages of GPU accelerated effects in scientific computing.

## 3.1 Matrix multiplication

Matrix multiplication has a high degree of parallelism. In this experiment, through different dimensions of matrix multiplication, we compare speedup ratios of different dimensions matrix multiplication operations, Matrix data is randomly generated by float number.

## 3.2 Fast Fourier transform

FFT is similar to matrix computations, it have a high degree of parallelism. In this experiment, through different dimensions of matrix-FFT calculations, we compare speedup ratios. Matrix data is randomly generated by float number.

## 3.3 Quick sort

Quick sort algorithm contains a lot of logic, loop and branching jump instruction. This experiment was by sorting arrays of different lengths, we compare the speed of operate control-flow instruction of CPU and GPU. Array data is randomly generated by float number.

## 3.4 Simulate encoding and decoding Hamming code in BSC channel

In this comprehensive experiment, we looking for most efficient computing methods in scientific computing. Coding process of Hamming codes and channel simulation is matrix operations and a small amount of logical instructions, decoding process of Hamming code has a lot of logical instructions. In this experiment, different amounts of information sequence encoded (7, 4) Hamming code, and transmitted through the BSC channel, then decoding. Information sequence is randomly generated by binary number.

# 4. Experiment results and performance analysis

In this paper, we use Windows 8.1 operating system as test platform, CPU is Intel Core I5-2430M, which has dual cores running on 2.4GHZ, 6GB RAM. GPU is Nvidia GeForce GT540M, which has 96 unified shaders running on 672Mhz, 1GB VRAM.

## 4.1 Matrix multiplication

In this experiment, test matrix's size is 50*50, 100*100, 200*200, 500*500 and 1000*1000. We recorded the CPU and GPU operation time for whole experiment, and

GPU time plus the time of data transport between memory and VRAM. Time as shown in table 1.

As can be seen from table 1, when the matrix size is 50*50, CPU operation time is 0.000127s, GPU operation time is 0.000201s, GPU operation does not accelerate the computing, but lag behind the traditional method of calculation approximately 0.63 times. When the matrix size is 100*100 and 200*200, GPU operation slightly faster than CPU, when the matrix size is 500*500 and 1000*1000, GPU operation get a good speedup ratio, the speedup ratio even up to 52.92. But when we consider the time of data transport between memory and VRAM, GPU is slower than CPU, data transfer spending a lot of time.

**Table 1.** Matrix multiplication time of CPU and GPU.

| Dimension | CPU operation time | GPU operation time | Speedup ratio | GPU operation time and data transport time |
| --- | --- | --- | --- | --- |
| 50*50 | 0.000127s | 0.000201s | 0.63 | 0.000808s |
| 100*100 | 0.000226s | 0.000216s | 1.04 | 0.001625s |
| 200*200 | 0.000939s | 0.000726s | 1.29 | 0.002621s |
| 500*500 | 0.009632s | 0.000929s | 10.36 | 0.018578s |
| 1000*1000 | 0.091341s | 0.001726s | 52.92 | 0.238119s |

### 4.2 Fast Fourier transform

In this experiment, size of test matrix is 1024*32, 1024*128, 1024*512, 1024*1024 and 2048*2048. We recorded the CPU operation time and GPU operation time for whole experiment, and GPU time plus the time of data transport between memory and VRAM. Time as shown in table 2.

**Table 2.** FFT operation time of CPU and GPU.

| Dimension | CPU operation time | GPU operation time | Speedup ratio | GPU operation time and data transport time |
| --- | --- | --- | --- | --- |
| 1024*32 | 0.000332s | 0.000803s | 0.41 | 0.001946s |
| 1024*128 | 0.001963s | 0.000884s | 2.22 | 0.006362s |
| 1024*512 | 0.004775s | 0.000990s | 4.82 | 0.015180s |
| 1024*1024 | 0.008323s | 0.000884s | 9.41 | 0.019112s |
| 2048*2048 | 0.033987s | 0.001271s | 26.74 | 0.077520s |

Test results of FFT and matrix operations are similar, when the matrix size is small, GPU doesn't have any advantages. But with the increased dimension of the data, GPU computing efficiency becomes higher and higher. When the matrix size is 2048*2048, the speedup ratio even up to 26.74. And data transfer also spending a lot of time.

From the experiment in 4.1 and 4.2, we can see that programs with high parallelism running in GPU can get a good speed, and with the increased dimension of the data, GPU computing efficiency becomes higher and higher. But data transport between memory and VRAM will cost a lot of time. So when we use GPU to compute, we need to reduce times of data transfer, and try to increase the size of the calculation to get a higher computational efficiency of GPU.

### 4.3 Quick sort

In this experiment, size of test arrays is 500, 1000, 5000, 10000, 100000, 1000000. We recorded the CPU operation time and GPU operation time for whole experiment, Time as shown in table 3.

According to the testing results, we can see for control flow instructions, GPU efficiency is much lower than the CPU, when the matrix size is small, GPU's speed less than 1% of CPU. Only when the dimension is very large, GPU computing speed is close to the CPU.

**Table 3.** Quicksort operation time of CPU and GPU.

| Dimension | CPU operation time | GPU operation time | Speedup ratio |
|---|---|---|---|
| 500 | 0.000042s | 0.007459s | 0.0056 |
| 1000 | 0.000066s | 0.006963s | 0.0094 |
| 5000 | 0.000348s | 0.006811s | 0.0511 |
| 10000 | 0.000767s | 0.007172s | 0.1069 |
| 100000 | 0.005189s | 0.008114s | 0.6395 |
| 1000000 | 0.082952s | 0.031115s | 2.67 |

From the experiment in 4.3, we can see that programs with control flow instruction running in GPU are much slower than CPU. So when we use GPU to compute, try to avoid loops and switches.

### 4.4 Simulate encoding and decoding Hamming code in BSC channel

In this experiment, size of information sequence is 10000, 15000, 20000, 50000 and 100000. We recorded the CPU and GPU coding and BSC channel simulation time, and CPU and GPU decoding time. Operation time as shown in table 3.

According to the testing results, Coding process of Hamming codes and channel simulation is matrix operations and a small amount of logical instructions, and in the case of large dimension, GPU's efficiency is better than CPU. But decoding process of Hamming code have a lot of logical instructions, and GPU's speed is much slower than CPU.

**Table 4.** Coding and decoding Hamming code in BSC channel time.

| Dimension | CPU coding and BSC channel simulation time | GPU coding and BSC channel simulation time | Speedup ratio | CPU decoding time | GPU decoding time |
| --- | --- | --- | --- | --- | --- |
| 10000 | 0.001670s | 0.003288s | 0.0507 | 0.200791s | 37.580352s |
| 15000 | 0.004784s | 0.005957s | 0.8031 | 0.268413s | 64.931636s |
| 20000 | 0.011825s | 0.006859s | 1.7240 | 1.323330s | 87.918839s |
| 50000 | 0.035251s | 0.008626s | 4.0866 | 2.490690s | 209.230584s |
| 100000 | 0.063902s | 0.013541s | 4.7191 | 4.530879s | 429.688456s |

From the experiment in 4.4, we can see if we use GPU to compute operations have high parallelism, and use CPU to execute branch jump and the loop instructions, we will achieve a good efficiency.

## 5. Conclusion

When we use the GPU to compute, operation with high parallelism will produce a very high computational efficiency. The larger the scale of operation will get a higher efficiency. When we use the GPU to compute, try to avoid loops and switches, and reduce times of data transfer between RAM and VRAM.

According to the characteristics of the GPU, in scientific computing, CPU/GPU collaborative computing in large-scale scientific computing will get a better efficiency.